\documentclass [12pt,preprint]{aastex}

\usepackage{epsfig}
\begin{document}
\voffset-1cm
\newcommand{\gsim}{\hbox{\rlap{$^>$}$_\sim$}}
\newcommand{\lsim}{\hbox{\rlap{$^<$}$_\sim$}}

\title{The Magnetic Field in Galaxies, Galaxy Clusters, \\ and the 
InterGalactic Space}

\author{Arnon Dar\altaffilmark{1} and A. De R\'ujula\altaffilmark{2}}

\altaffiltext{1}{
Physics Department and Space Research Institute, Technion, Haifa 32000,
Israel.\\
arnon@physics.technion.ac.il, dar@cern.ch}
\altaffiltext{2}{Theory Division, CERN, 1211 Geneva 23, Switzerland,
and Physics Dept.~Boston University, USA.
alvaro.derujula@cern.ch}

\begin{abstract}

Magnetic fields of debated origin appear to permeate the Universe on all
large scales. There is mounting evidence that supernovae produce not only
roughly spherical ejecta and winds, but also highly relativistic jets of
ordinary matter. These jets, which travel long distances, slow down by
accelerating the matter encountered on their path to cosmic-ray energies.
We show that, if the turbulent motions induced by the winds and the cosmic
rays generate magnetic fields in rough energy equipartition, the predicted
magnetic-field strengths coincide with the ones observed not only in
galaxies (5 $\mu$G in the Milky Way) but also in galaxy clusters (6 $\mu$G
in Coma). The prediction for the intergalactic (or inter-cluster)  field
is 50 nG.

\end{abstract}

\section{Introduction}

The average magnetic field (MF) of the interstellar medium (ISM) of our
Galaxy ($B_{_{\rm MW}}\!\sim\! 5\,\mu\rm{G}$)  corresponds to an energy
density of $\sim 0.5$ eV cm$^{-3}$, in good agreement with the energy
density of cosmic rays (CRs). This provides a strong hint of a common
origin and of energy equipartition (see, e.g.~Longair 1992), though other
theories of the origin of galactic fields ---e.g. by dynamo amplification
of primordial seed fields--- have been proposed (Parker 1992). The origin
of intergalactic MFs, both within and without galaxy clusters, is also
undecided. Here we discuss how all of these MFs could have a common
origin, and be in equipartition with the corresponding local energy
densities of CRs.
 
Radio observations of clusters indicate that the {\it intra}-cluster
medium (ICM) between the galaxies is permeated by intense MFs
(e.g.~Kronberg 2004). Nearby clusters are seen to have a ``radio halo"  with a
distribution similar to that of the cluster gas, observed in X-rays. These
halos are produced by synchrotron emission from CR electrons spiralling in
the cluster's MF, while the X-rays are electron bremsstrahlung.
Measurements of the Faraday rotation of linearly polarized radio emission
traversing the cluster's medium, in combination with X-ray data, support
the existence of cluster MFs of a few $\mu$G (Kim et al. 1989, 1990; Taylor
\& Perley 1993; Feretti et al. 1995; Deiss et al. 1997; Eilek 1999;  
Ensslin et al. 1999; Clarke, Kronberg \& Bohringer 2001;  Johnston-Hollitt, Hollitt
\& Ekers 2004). The mapping of the Faraday rotation reveals that the
clusters' MFs are turbulent with a Kolmogoroff power spectrum on a variety
of scales (Ensslin, 2004; Vogt \& Ensslin 2004).

The MF between clusters and isolated galaxies in the {\it inter}-galactic
medium (IGM) is not known. Speculations on its value range from nearly a
$\mu$G to a pG. Low-level radio emission was detected from the IGM
around Coma (Kim et al.~1989; Ensslin et al.~1999) and from the IGM in 
large-scale filaments of galaxies (Bagchi et al.~2002). The estimated
field strengths are of the order of several hundred nG.

Theories of the origin of MFs in the ICM and IGM include cosmic shocks 
(e.g.~Kulsrud et al.~1997; Ryu, Kang \& Biermann 1998), ionization fronts (Gnedin 
et al.~2000) and outflows from primeval galaxies (Kronberg et al.~2004), quasars and/or
radio galaxies (Furlanetto \& Loeb 2001, Gopal-Krishna \& Wiita 2001).
Kronberg et al.~(2004) have estimated that ``giant" extragalactic radio sources, 
powered by  accretion
onto massive black holes ($M>10^8\, M_\odot$), inject $E_B\sim 10^{60-61}$ ergs of
magnetic energy into radio lobes, and have argued that the expansion and
diffusion of these Mpc-scale lobes could have magnetized a
large fraction of the IGM. Assuming that in the accretion $\sim\!1\%$ of $M$
is released in the form of magnetic energy,
Kronberg (2004) estimated a mean $B_{_{\rm IGM}}\!\sim\! 40$ nG at redshift $z\!\sim\!2$.
This value\footnote{If the ratio of the mass of the central
massive black hole and the luminosity of the Galaxy
($\sim\!1.5\times 10^{-4}\, M_\odot/L_\odot$) is universal,
the luminosity density of the local universe 
($\sim\!1.2\times 10^8\, L_\odot$ Mpc$^{-3}$) implies a black hole mass
density of $\rho_{_{\rm BH}}(z=0)\!\sim\!1.8\times 10^4\, M_\odot$ Mpc$^{-3}$,
and $\rho_{_{\rm BH}}(z=2)\!\sim\!1.5\times 10^5\, M_\odot$
Mpc$^{-3}$ at the ``quasar epoch'', consistent with
$\rho_{_{\rm BH}}(z=2)\!\sim\!2.2\times 10^5\, M_\odot$
 Mpc$^{-3}$, adopted by
Kronberg (2004).} evolves as $(1+z)^{-2}$ by cosmic expansion, yielding a 
two-orders-of-magnitude smaller IGM energy density, and a one-order-of-magnitude
smaller $B_{_{\rm IGM}}$ at $z\!=\!0$.

Concerning the ICM, it was suggested that the jets formed by accretion
onto massive black holes in clusters provide the heat source in the 
so-called ``cooling flow" (CF) clusters (e.g.~McNamara et al.~2000). But
Kronberg et al.~(2004) have also found that the radio lobes of the
powerful radio galaxies at the centre of rich clusters contain a magnetic
energy of only $E_B\sim 10^{58-59}$ erg.  Assuming equipartition between
the kinetic energy output and the magnetic energy, the energy supply from
such objects is insufficient to power the X-ray emission from bright CF
clusters over a Hubble time ($\sim 10^{62}$ erg for bright CF clusters).
Moreover, some CF clusters contain neither powerful radio galaxies nor
active galactic nuclei. Contrariwise, Colafrancesco, Dar \& De R\'ujula
(2004) have shown that the required heat supply in CF clusters can be
provided by the energy deposited in the ICM by jets and CRs emitted from
the cluster galaxies\footnote{For the CRs to be the heating agent of the
cluster gas, the CR luminosity of a Galaxy must be a few times its
``classic" estimate (Colafrancesco et al. 2004). For the related
interpretation of the Gamma Background Radiation to
succeed, the jets must travel a few kpc or more (Dar \& De R\'ujula 2001).
Both requirements are predictions of the ``Cannonball" model of
high-energy astrophysics (e.g.~De R\'ujula 2004a,b).}, 
and that the equipartition of energy between
the ionized gas, the MF and the CRs can explain the origin of MFs of
several $\mu$G.

In this letter we argue that the outflow of jets, CRs and winds from SN
explosions in star formation regions, where most SNe take place, magnetize
the ISM in galaxies, the ICM in galaxy clusters and the IGM outside galaxy
clusters. Our main assumption is that of a rough equipartition between the
energy of the SN jets and winds and the energy of the accompanying CRs.
The predicted strengths of the MF in the ISM and ICM are consistent with
those observed. In large structures, the predicted magnetic energy density
is roughly proportional to the luminosity density, with a mean MF of
several $\mu$G in the ICM of rich galaxy clusters and $\sim\!50$ nG in the
IGM.

\section{Galactic CRs, magnetic fields and supernova explosions}

As a result of the steep energy spectrum of galactic CRs, the bulk of the 
CR energy is carried by nuclei with an average energy of a few GeV.
The most accurate  measurements of their flux, $dI/dE$, near Earth and
during solar minimum (minimum solar modulation) are those of
AMS (Alcaraz et al.~2000a,b) and BESS (Haino et al.~2004). 
Their measurements yield a local CR energy density\footnote{At  energies
below $\sim\! 1$ GeV,
$dI/dE$ is affected by Earth and Sun's effects. The uncertainties on the
energy-weighed  integral of Eq.~(\ref{rhoECR}) are much smaller.}:
\begin{equation}
\rho_{_{E}}{\rm [CR]}={4\,\pi\over c}\,\int {dI\over dE}\, E\,  dE\approx 0.5\, 
                \rm eV\, cm^{-3}.
\label{rhoECR}
\end{equation}
If the energy densities of galactic CRs and 
MFs are in equipartition, 
\begin{equation}
{B_{\rm MW}^2/ (8\, \pi)} \approx \rho_{_{E}}\rm[CR],
\label{rhoEB}
\end{equation}
and $B_{\rm MW}\sim 5\, \mu$G,
in good agreement with observations, as is well known (Longair 1992).

There is evidence from gamma-ray bursts (e.g.~Dar \& De R\'ujula 
2004, Dar 2004a), from SN1987A (Nisenson \&
Papaliolios 1999), and from the morphology of young supernova (SN) remnants
(e.g.~Hwang et al.~2004) that in addition to quasi-spherical ejecta, SNe
produce highly relativistic and narrowly collimated jets,
which carry an average kinetic energy $E_K\rm[Jet]\!\sim\! 2\!\times\! 
10^{51}$
erg, similar to the kinetic energy of the spherical ejecta. The jets slow down
by collisions with the interstellar matter (ISM) along their path.
The intercepted ISM is thereby accelerated to CR energies, carrying away
almost entirely the original energy of the jets. This simple theory of CRs
agrees very well with the observed CR spectra and CR composition at all energies
(Dar 2004b; De R\'ujula 2004a,b). The fast winds also transport CRs and
magnetic fields within galaxies and out of them. We assume equipartition in the
sense that, of the total energy ($\sim\!2\,E_K\rm[Jet]$) of winds and jets, 1/2 ends 
up in CRs.

Can the bulk of the galactic CRs be accelerated by SNe in the stated way?
If SNe produce the observed flux  of galactic CRs, 
the galactic CR luminosity must satisfy:
\begin{equation}
L_{\rm CR}{\rm [MW]}\approx
{4\, \pi\, V_{\rm CR}{\rm [MW]}\over c}\, \int {dI\over dE}\, 
{E\over\tau(E)}\,dE\approx 
R_{\rm SN}{\rm [MW]}\, E_K{\rm[Jet]}\approx 1.3\times 10^{42}\,\rm erg\, s^{-1}.
\label{SNEsupply}
\end{equation}
where  $V_{\rm CR}\rm[MW]$ is the confinement volume 
of low-energy CRs in the Galaxy, $\tau(E)$ is their mean
confinement time, and $R_{\rm SN}\rm[MW]$ is the present Galactic SN rate.
In the numerical result we have used the estimate
$R_{\rm SN}\rm[MW]\!\approx\! 1/50$ y$^{-1}$, obtained from the rate and
spatial distribution of historical SNe and the measured galactic
extinction, and the quoted value of  $E_K\rm[Jet]$.  
 This SN rate is also consistent with the value measured 
in the local universe\footnote{The SN rate is  2.8 y$^{-1}$ (Van den Bergh \& 
Tammann, 1991) in a ``fiducial sample" of 342 galaxies within the
Virgo circle (whose total B-band luminosity is 
$1.35\,h^{-2}\times 10^{12}\,L_\odot^B$). 
For $h=0.65$ ($H_0$ in units of 100
km Mpc$^{-1}$ s$^{-1}$)   and the galactic luminosity
$L_{\rm MW}\!\sim\! 2.4 \times 10^{10}\, L_\odot $, we also obtain 
$R_{\rm SN}\rm[MW]\!\approx\! 1/50$ y$^{-1}$.}. 

By fitting the diffuse gamma-ray emission of the Galaxy, as measured by
EGRET (Sreekumar et al.~1998), to the CR production rate of $\gamma$-rays 
in a ``leaky box" model of
the galactic CR halo, Strong et al.~(2004) obtained 
 $V_{\rm CR}\rm[MW]\!\approx\! 2.1\times 10^{68}$ cm$^{3}$ 
 (the volume of a cylinder with 30
kpc diameter and 10 kpc height). Using the observed CR spectrum
and $\tau(E) \!\sim\! 2\times 10^7\,
(E/\rm GeV)^{-0.5\pm 0.15}$  y (inferred from the relative
abundance of unstable isotopes in CRs), we obtain for the mean
injection rate of CR energy per unit volume in the Milky Way 
$\dot{\rho}_{_{E}}\rm [CR]\!\approx\! 1.8\times 10^{-19}$ erg cm$^{-3}$
y$^{-1}$, and consequently $L_{\rm CR}\rm[MW]\!\sim\! 1.2\times 10^{42}$ erg
s$^{-1}$. This number  agrees with the RHS of  Eq.~(\ref{SNEsupply}).

From the above we conclude that SN explosions seem to be 
the source of the bulk of the CR  and  MF
energies in the ISM of ordinary galaxies.  

\section{The magnetic field in the ICM of galaxy clusters}

Let $R(z)$ be the rate of SN events in a galaxy such as ours, at
redshift $z$, or look-back time $t$. For a standard cosmology 
with $\Omega=\Omega_M+\Omega_\Lambda=1$,
$dt/dz= [\Omega_M\, (1+z)^3 +\Omega_\Lambda]^{1/2}/[H_0\, (1+z)]$,
where $\Omega_M\!\approx\! 0.27$, $\Omega_\Lambda\!\approx\!0.73$ and
$H_0\!\approx\! 65$ km Mpc$^{-1}$ s$^{-1}$.
The SN rate is proportional
to the star formation rate $R_{\rm SF}(z)$, so that 
$R(z)=R(0)\, R_{\rm SF}(z)/R_{\rm SF}(0)$. The observations
(see, e.g.~Lilly et al.~1995; Madau et al.~1996, Steidel et al.~1999;
Schiminovich et al.~2004) are:
$R(z)\!\sim\! R(0)\, (1+z)^\alpha$ with 
$\alpha\!\approx\! 2.5\pm 0.5$ for $z\leq 1$ and
$R(z)\approx R(1)\,([1+z]/2)^{0\pm 0.5}$ for 
$1\leq z\leq 5$, a redshift beyond which the relative volume is
 small. If the star formation history in a galaxy
cluster (GC) is not very different from that in the Milky Way
or the rest of the universe,  the cluster's SN rate is simply
weighed by the ratio of luminosities:
$R_{\rm GC}(z)\!\approx\! (L_{\rm GC}/L_{\rm MW})\, R(z)$. 
For reasonable MF coherence
lengths, low-energy CRs do not diffuse out of 
a rich cluster during a Hubble time, and the total CR energy in
their ICM is:
\begin{equation}
E_{\rm CR}(z_o)= {2\over 3}\,E_K{\rm [Jet]}\,{R_{\rm SN}{\rm [GC]}\over H_0}\, 
                    \int_{z_o}^\infty \,
                  {R_{\rm SF}(z) \over R_{\rm SF}(0)}\, {dz\over (1+z)
                  \sqrt{\Omega_M\, (1+z)^3 +\Omega_\Lambda}}\, ,
\label{EzGC} 
\end{equation}
if the cluster decouples
from the Hubble expansion at a relatively early time. The factor  2 
reflects the equality of energies of jetted and ``spherical" ejecta; the factor 1/3 the
energy equipartition between CRs, MFs, and the {\it dense}\footnote{CR-induced
hadronic and electromagnetic showers are efficient in transferring energy
to the dense ICM plasma, but not to the thin ISM or IGM  (Colafrancesco
et al.~2004).} ICM plasma.

For $z_o=0$, the integral in Eq.~(\ref{EzGC}) is
$\approx\!2.6\pm 1.3$, implying a 
CR energy density:
\begin{equation}
\rho_{\rm ECR}{\rm [ICM]}\sim {L_{\rm GC}\over L_{\rm MW}}\, 
{ R_{\rm SN} {\rm [MW]}\, E_K{\rm [Jet]} 
\over H_0\, V_{\rm GC}}
\sim (0.22\pm 0.11)\,\left({L_{\rm GC}\over 10^{12}\,L_\odot}\right)\,
\left({{\rm Mpc} \over R_{\rm GC}}\right)^3\,\rm  eV\, cm^{-3}.
\label{rhoeGC}
\end{equation}
If the  CRs from SN explosions magnetize 
the ICM in the same way that they magnetize the ISM, the 
MF in the ICM has the same energy density as the CRs:   
$B_{\rm ICM}^2 / (8\, \pi)\!\sim\! \rho_{\rm ECR}$.
This prediction is in good agreement with observations. For instance,
the observed luminosity of Coma  within a radius of 1 Mpc
is $L_{\rm Coma}\!\approx\! 3\times 10^{12}\, L_\odot$ 
(Fusco Femiano \&
Hughes 1994), implying a CR energy density, 
$\rho_E{\rm [CR]}\!\sim\! (0.67\pm 0.33)$ eV cm$^{-3}$,
and $B_{\rm Coma}\!\approx\! 5.1\pm 1.2 \,\mu$G, in 
agreement with the observed $\sim\!6\, \mu$G (e.g.~Clarke et al. 2001).

\section{Intergalactic cosmic rays and magnetic fields}

Let $R_{\rm SN}(z=0)\!\approx\! 10^{-4}$ Mpc$^{-3}$ y$^{-1}$
be the current local SN rate per comoving unit volume
(the observed SN rate per unit luminosity within the 
Virgo circle, multiplied by $\rho_L\approx 1.2\times 10^8\, L_\odot$
Mpc$^{-3}$, the mean luminosity density in the local universe).  
In a steady state, the injection rate of CRs into the 
IGM by a galaxy is equal to its CR production rate. 
Consequently, SN explosions in galaxies, at a cosmic time $t(z)$,
inject energy into the IGM at a rate $\sim 2\, E_K{\rm [Jet]} \,R_{\rm SN}(z)$.
Let $dN_{\rm SN}/dE$ be the CR spectrum produced by a single SN.
Its energy dependence is that of $(dI/dE)/\tau(E)$, with $dI/dE$
the observed CR spectrum, and $\tau(E)$ the CR the residence time. 
In equipartition, its normalization is $\int (dN_{\rm SN}/dE)\,E\, 
dE=E_K{\rm [Jet]}$.
The CR spectral density in the IGM at redshift $ z_o $, is given by:
\begin{equation}
{dn\over dE}(z_o)= {R_{\rm SN}(0)\over H_0}\, \int_{z_o}^\infty \, 
{dz\over\sqrt{\Omega_M\, (1+z)^3 +\Omega_\Lambda}}\;
                  {R_{\rm SF}(z) \over R_{\rm SF}(0)}\, 
                  {dN_{\rm SN}\over dE'}\Bigg\vert_{E'=(1+z)\,E}\; .
                  \label{dndezIGM}
\end{equation}  
The present CR energy density of the IGM implied by equipartition and
Eq.~(\ref{dndezIGM}) is:    
\begin{equation}
\rho_{\rm ECR}{\rm [IGM]}=\int {dn\over dE}\, E\, dE\approx (0.76\pm 0.38)\times 
10^{-4}\, \rm eV\, cm^{-3}. 
\label{dndeIGM}
\end{equation}
If the galactic winds and CRs from SN explosions magnetize 
the IGM in the same way that they magnetize the ISM, then, 
under the assumption of equipartition, the 
magnetic field in the IGM 
has the same energy density as the CRs:     
$B_{\rm IGM}^2 / (8\, \pi)=\rho_{\rm ECR}\rm [IGM]$. Hence
the  average strength of  magnetic fields in the IGM is predicted to 
be $B_{\rm IGM}\!\sim\! 50\pm 12$ nG. The estimated field in the outskirts
of Coma (several hundred nG; see Kim et al.~1989; Ensslin et al.~1999) 
is intermediate between
our expectations for the ICM and the IGM.

\section{Conclusions} There is evidence that long-duration gamma ray
bursts are produced by relativistic jets ejected in core-collapse SN
explosions, as long advocated in the ``Cannonball" (CB) model (Dar \& De
R\'ujula 2004; Dar 2004a and references therein). The jets, along their
long paths (much larger than a galaxy's size), transfer essentially
all of their energy to the local medium, which is accelerated to CR
energies (Dar 2004; De R\'ujula 2004, Dar \& De R\'ujula in preparation).  
The generation of CRs and the subsequent MFs along the jet's path is fast:
it occurs at nearly the speed of light. It is known from
``first-principle" simulations that a relativistic plasma (in our case,
the CRs) impinging on a medium at rest generates turbulence and MFs
 efficiently and very rapidly (Frederiksen et al.~2004). The transport of CRs and
MFs by SN winds, even at a modest few thousand km s$^{-1}$ reaches ---in a
Hubble time--- distances larger than the mean separation between galaxies.
This justifies our implicit assumption that, in equipartition with CRs,
sufficiently uniform MFs are generated
in a time much shorter than Hubble's time.

In equilibrium, the CRs escaping a galaxy ---or being generated by jets
beyond the galaxy's confines--- have the same spectrum: the CR ``source
spectrum": the observed  spectrum deprived of the galactic
confinement-time effect, as in Eq.~(\ref{SNEsupply}). We have assumed
equipartition between CR and MF energies in the low-density ISM and IGM,
and between the energies of CRs, MFs and the dense plasma of the hot
central regions of the ICM. On this basis, we obtained $B_{\rm MW}\sim 5\,
\mu$G for the mean magnetic field in the Galaxy, and a very similar
value for $B_{\rm Coma}$, the mean magnetic field in the ICM within 1 Mpc
from the centre of the Coma cluster (or similarly rich clusters), in
agreement with the observations. The prediction for the IGM is
$B_{\rm IGM}\sim 50$ nG. The observations in the outskirts of Coma are between
our predictions for the ICM and the IGM, but not enough is known about the
propagation of CRs in the IGM to claim that this is a success.

Our theory of large-scale MFs also explains in a very simple fashion the
properties of CRs (Dar 2004; De R\'ujula 2004). It individuates the
heat-source in ``cooling flow'' clusters and predicts the temperature
profile of the intra-cluster gas (Colafrancesco et al.~2004).  It predicts
an extragalactic $\gamma$-ray background radiation with a spectral index
$\sim\!-2.1$, dominated by inverse Compton scattering (ICS) of the
microwave background radiation by CR electrons in the galactic halo and in
the IGM. A similar radiation from the halo of Andromeda may be observable
by GLAST (Dar \& De R\'ujula 2001). The theory entails a $\gamma$-ray
emission from the ICM of clusters of galaxies, due to ICS of CR electrons
and $\pi^0$ production and decay (with spectral indices $\sim\!-2.1$ and
$\sim\!-2.2$, respectively)  in the collisions of CR nuclei with the ICM
(Dar \& De R\'ujula 2001). These radiations are also at a level detectable
by GLAST.

Although other accelerators ---such as flaring stars, stellar winds,
SN remnants, pulsars, microquasars, and massive black holes in
active galactic nuclei--- contribute to the production of non-solar
CRs, supernova explosions seem to be the dominant source of CRs, 
as speculated long ago (Baade \& Zwicky 1934). 
Not only can the SN outflows accelerate the bulk of the high
energy CRs, but they can also magnetize the entire universe at the
observed level.

\noindent 
{\bf Acknowledgements:} The autors would like to thank Etienne
Parizot for useful comments. A. Dar is grateful for the hospitality of the
Theory Unit at CERN. This research was supported in part by the Helen
Asher Fund for Space Research at the Technion.

\end{document}